# A TEM study of morphological and structural degradation phenomena in LiFePO$_4$-CB cathodes

Duc-The Ngo,[1,2(*)] Roberto Scipioni,[1] Søren Bredmose Simonsen,[1] Peter Stanley Jørgensen,[1] and Søren Højgaard Jensen[1(**)]

[1] DTU Energy, Department of Energy Conversion and Storage, Technical University of Denmark, Frederiksborgvej 399, 4000 Roskilde, Denmark

[2] Electron Microscopy Centre, School of Materials, University of Manchester, Oxford Road, Manchester M13 9PL, United Kingdom

**Abstract**

LiFePO$_4$-based cathodes suffer from various degradation mechanisms which influences the battery performance. In this paper morphological and structural degradation phenomena in laboratory cathodes made of LiFePO$_4$ (LFP) mixed with carbon black (CB) in an 1 mol L$^{-1}$ LiPF$_6$ in EC:DMC (1:1 by weight) electrolyte are investigated by transmission electron microscopy (TEM) at various preparation, assembling, storage and cycling stages. High-resolution TEM (HRTEM) imaging shows that continuous SEI layers are formed on the LFP particles and that both storage and cycling affects the formation. Additionally loss of CB crystallinity, CB aggregation and agglomeration is observed. Charge-discharge curves and impedance spectra measured during cycling confirm that these degradation mechanisms reduce the cathode conductivity and capacity.

**Corresponding author**: (*) duc-the.ngo@manchester.ac.uk (Duc-The Ngo), (**) shjj@dtu.dk (Søren Højgaard Jensen)

## 1. Introduction

Lithium iron phosphate, LiFePO$_4$ (LFP) is a common cathode material in lithium-ion batteries.[1,2] It combines reasonably good cycle life, low cost and toxicity, high Open Circuit Voltage (OCV) around 3.4 V vs. Li$^+$/Li and a theoretical charge/discharge capacity of ~170 mAh/g.[3–5] LFP has a low electronic conductivity which requires mixing with carbon or a conductive polymer to ensure good electric conductivity and power density.[6,7,8]

Several degradation phenomena can occur in this type of cathode. Among these, cycling-related micro-cracks in larger LFP grains,[9–11] loss of crystallinity[12] and active material[9] as well as carbon aggregation and agglomeration,[13,14] are known to negatively affect the electrode performance and cause capacity fading. LFP nanoparticles are surface sensitive to modifying additives.[15–19] Both the electrolyte composition, the use of additives and the electrode charge/discharge history affects the morphology of decomposition compounds, the SEI layer formation, and the related electrode performance.[20–23]

Here we present a study of laboratory LFP electrodes in 1 mol L$^{-1}$ LiPF$_6$ in EC:DMC (1:1 by weight) at different preparation, assembling and testing stages. TEM microscopy of as-received LFP and CB powders, as-prepared LFP-CB, LFP-CB stored in the electrolyte and LFP-CB after 100 charge/discharge cycles is used to investigate nano-structural changes including SEI layer formation. A heterogeneous electrode structure, formation of secondary phases and multi-layered SEI is observed in the cycled electrode. Although not investigated in detail in this paper, the latter interestingly indicate that the electrode history - to some extent - is stored and can be detected in the SEI layers.

## 2. Experimental Details

Laboratory LFP-CB cathode specimens were prepared from LFP nanoparticles (MTI Corp., US), Super C65 carbon black (Timcal, Switzerland) and polyvinylidene fluoride (PVdF). The microstructure of as-received LFP and CB nanoparticles were studied separately and referred to as "pristine" samples. The LFP, CB and PVdF were mixed with a ratio of 80:10:10 and dissolved in N-Methyl-2-pyrrolidone (NMP) solvent by magnetic stirring for 10 hours. The PVdF was used as binder to enhance the adhesion to the current collector. After magnetic stirring, a TEM specimen was prepared by putting a small drop of the cathode mixture on Au TEM grids with a holey carbon support film. Subsequently it was dried at 120$^{o}$C under vacuum. The sample was investigated in the TEM microscope and is referred to as the "fresh cathode". After TEM characterization, the fresh cathode was kept in a standard 1M LiPF$_6$ in 1:1 EC/DMC electrolyte for 72 h, rinsed with diethyl carbonate and dried at 120$^{o}$C under vacuum. The sample was investigated once again in the TEM and referred to as the "stored cathode".

Two electrodes were prepared by drying the LFP, CB and PVdF solution in NMP solvent (described above) on an Al current collector at 120$^{o}$C under vacuum. The electrodes were then subjected to respectively 2 and 100 charge/discharge cycles in a three-electrode EL-CELL® ECC-Combi cell house at 0.1 C using lithium metal foil as counter and reference electrodes and a glass fiber separator soaked with the standard 1M LiPF$_6$ in 1:1 EC/DMC electrolyte. After cycling, the cell houses were disassembled in a glove box, and the two electrodes were rinsed with diethyl carbonate to remove the remaining electrolyte before being dried at 120$^{o}$C under vacuum and subsequently embedded in silicon resin (Wacker Chemie). The electrodes cycled 2 times and 100 times are referred to as the "reference cathode" and "aged cathode", respectively. After curing of the resin, a TEM lamellar specimen of the aged cathode was prepared by focused ion beam (FIB) milling using a 30 kV Ga ion beam (Zeiss Crossbeam XB1540). Scanning Electron Microscopy (SEM) images of the reference and aged cathodes were also obtained using FIB/SEM imaging with the same Crossbeam XB1540 equipment. An overview of the studied samples is presented in Table 1.

(S)TEM imaging (bright-field, high resolution and high annular angle dark field imaging) of the LFP-CB cathode specimens was performed on a JEOL JEM 3000F equipped with a 300 kV field emission gun (FEG), high annular angle dark field (HAADF) STEM detector, and an Oxford Instruments X-ray detector with an ultra-thin window for X-ray energy dispersive spectroscopy (EDX) analysis.

Electrochemical impedance spectroscopy (EIS) measurements were performed on the two cycled cathodes in the three-electrode EL-CELL® ECC-Combi using lithium metal as counter and reference electrodes. EIS measurements in a frequency range from 10 mHz to 10 kHz were obtained from the LFP-CB reference and aged electrodes in the discharged state at OCV, after the cells had reached a steady state defined by a voltage change less than 5 mV/h.

## 3. Results

*3.1 SEI Layers*

Fig. 1(a) displays a bright- and a dark-field (inset) TEM image of pristine off-the-shelf LFP nanoparticles deposited on a lacey carbon grid. A high-resolution TEM (HRTEM) image of one of the LFP particles is presented in Fig. 1(b). A thin amorphous coating with a thickness of ~1 nm is observed at the particle surface. Fig. 1(c) and 1(d) shows respectively a bright-field TEM and a HRTEM image of pristine CB nanoparticles. The inset in Fig. 1(c) shows a selected area electron diffraction (SAED) pattern of the CB particles.

Fig. 2(a-c) presents bright-field TEM images of the (a) fresh, (b) stored and (c) aged cathode microstructures with LFP (dark contrast, large particles) and CB nanoparticles (light contrast, smaller particles) distinguished by amplitude contrast.[24] In Fig. 2(c), the relatively dark LFP particles are indicated by black arrows and the lighter CB particles are indicated by white arrows. In Fig. 2(c) the grey contrast background arises from the Si resin stabilizing the TEM lamella.

Changes in the coating layer at the LPF particles surfaces are observed in the magnified TEM images [Fig. 2(d-f)]. It is seen that the thickness of the layers formed at the LFP particle surface increases from the fresh [Fig. 2(d)] to the stored [Fig. 2(e)] and further to the aged [Fig. 2(f)] cathode. Representative HRTEM images are shown for the fresh [Fig. 2(g)], stored [Fig. 2(h)] and aged [Fig. 2(i)] cathode. The thickness of the amorphous layer at the LFP particle surfaces increases from ~3 nm for the fresh cathode [Fig. 2(g)] to ~9 nm for the stored cathode [Fig. 2(h)] and ~30 nm for the aged cathode [Fig. 2(i)]. In the latter sample the SEI consist of distinct layers. The relation to the cathode history is debated further in the discussion section. The inset in Fig. 2 (g) shows a magnification of one of the LFP/SEI layer interfaces for the fresh sample. Further magnification of one of the LFP/SEI interfaces for the aged cathode sample can be seen in the supplementary data, Fig. S1(a). Fourier transforms of areas within the LFP particle and coating layer is presented in S1(b) and (c), respectively. Area selections for the Fourier transforms are shown with dotted squares in Fig. S1(a).

*3.2 Carbon Crystallinity*

SAED diffraction patterns of CB from the fresh, stored and aged samples are presented in Fig. 3(a), (b) and (c), respectively. The diffraction patterns were obtained from the areas inside the dotted blue rings in Fig. 2(a), Fig. 2(b) and Fig. 2(c) respectively. Diffraction intensity profiles are shown as insets in the figure. The diffraction intensity profiles are obtained by rotationally averaging the diffraction patterns and normalizing. A quantitative analysis of the peaks in the diffraction profile is provided in Table 2.

*3.3 Carbon agglomeration*

A STEM-HAADF image of the aged sample is presented in supplementary data [S2(a)]. Elemental maps of the sample using STEM-EDX are presented in Fig S2(b)-(f) and an average EDX spectrum showing apparent presence of C, O, Fe, P, Al and Si is illustrated in Fig. S2(g). It is noteworthy that Fluorine (F) is presumably present in the sample denoted by a shoulder (~677 eV) nearby the Fe peak (704 eV). However, it is likely impossible to resolve the F-presence because of small difference between Fe and F peaks (27 eV) compared to measurement resolution of our EDX detector (~140 eV). Fe is known to be part of LFP and thus present in the sample firmly, whereas it cannot be concluded from the EDX map whether F is present in the sample. Additionally the Al map is not provided since the Al counts were too weak to form a clear visible spatial distribution.

In the STEM HAADF image carbon appears as dark contrast areas, whereas the Si-epoxy resin is visible as grey contrast and the LFP is bright-contrast, in accordance with the Z-contrast rule in HAADF imaging. A high-magnification BF-TEM image of the carbon as seen in the upper part of Fig. S2, is presented in Fig. S3(a) and a STEM-HAADF image of the same region is shown in Fig. S3(b).

SEM images of the reference and aged electrode are presented in supplementary data [Fig. S4]. Relative to the reference electrode increased heterogeneity and formation of agglomerates is observed in the aged electrode.

*3.4 Charge-Discharge Cycling*

Fig. 4 displays 0.1 C charge-discharge curves for the reference and aged LFP-CB cathode. The two cathodes show similar initial charge capacity, but the aged cathode shows a significant reduction of charge capacity during cycling, ending at ~ 40 mAh/g after 100 cycles. The DC resistance – here calculated as the difference between the horizontal voltage level during charging and discharging (i.e. two times a DC overvoltage of 37 mV), divided with two times the applied DC current current – is fairly constant with cycling around 2000 $\Omega\ cm^2$.

*3.5 Impedance Analysis*

Fig. 5 shows impedance spectra recorded for the reference cathode and for the aged cathode at various state of charge (SOC) and cycling number. The equiaxial Nyquist plot, Fig. 5(a), shows impedance spectra for the aged cathode recorded at decreasing SOC from 100% to 0% SOC after 5 full cycles at 0.1 C. Fig. 5(b) shows a Bode plot of the same spectra. A Nyquist plot of impedance spectra for both the reference and aged cathode at various cycling number is presented in Fig 5(c). The spectra were recorded at 0% SOC, i.e. at 3V. A Bode plot of the same spectra is presented in Fig. 5(d). Zooms of the spectra are presented in Fig. 5(e) and (f).

## 4. Discussion

*4.1 SEI Layer formation*

From Fig. 1(b) it is seen that the pristine LFP particle is crystalline and has a thin amorphous coating with a thickness of ~1 nm at the surface. The coating is an electron conductive amorphous carbon layer deposited by the supplier.

The relatively distinct thin circles in the inset in Fig. 1(c) shows that CB is quasi crystalline in agreement with previously findings.[25,26] Fig. 1(d) reveals concentric single-crystal sheets in CB particles oriented approximately parallel to the electron beam. A distance of 0.34±0.02 nm between the sheets was obtained from a Fourier transform of the HRTEM image. This is in agreement with the previously reported distance of 0.36 nm between CB(002) planes.[26–28]

The overall contrast and morphology of the aged cathode [Fig.2(c)] seems to be different from that of the fresh [Fig. 2(a)] and stored [Fig. 2(b)] cathodes. However, at this magnification the main difference in appearance is due to the difference in sample preparation. The fresh and stored cathode samples were prepared on Lacey carbon grids whereas the aged cathode sample was a TEM lamella from the cycled cathode. In the lamella [Fig. 2(c)], the grey contrast background arises from the Si resin filling the pores between the LFP and CB particles while stabilizing the TEM lamella.

The measured difference in the thickness of the amorphous layer on the surface of the LFP particles in the pristine (~1 nm) and fresh sample (~3 nm) [Fig. 1(b) and Fig. 2(g)] could possibly be ascribed to the electrode preparation,[29] however it should be noted that the difference is comparable to the measurement uncertainty.

The Fourier transform of the LFP coating layer in [Fig. S1(c)] shows broad diffuse rings thus confirming that the coating is amorphous. The Fourier transform of the LFP particle in [S1(b)] shows bright spots. The measured lattice spacing's agree with the distances between LFP lattice planes within an estimated measuring error of 6%. This confirms that the crystalline structure of the LFP particle survived the cathode preparation and subsequent cycling.

The composition of the coating on the pristine LFP particles was not provided by the manufacturer. The EDX Al signal was too weak to provide a map with enough spatial resolution, however from the EDX spectrum [Fig. S2(g)] Al is seen to be present in the aged sample. This indicates the coating on the pristine LFP particles could be an Al-containing coating such as $AlF_3$ which are known to suppress iron dissolution and improve the cycling capability.[30,31]

It is observed that while the coating layers in the fresh [Fig. 2(g)] and stored [Fig. 2(h)] cathodes appear as one single amorphous layer, the coating layer in the aged cathode [Fig. 2(i)] seems to consist of several layers distinguishable by different contrast levels.

The observation of a growing amorphous SEI layers on LFP during exposure to the electrolyte agrees with previous reports.[32,33] It has also been reported that SEI layers can grow during battery charge/discharge cycling and dismantling.[33,34]

Specifically, LFP nanoparticles are surface sensitive to modifying additives which means the electrolyte composition is important for the morphology of decomposition compounds and SEI layers formed on $LiFePO_4$ surfaces.[15–18] For instance, continuous SEI layer formation on $LiFePO_4$ using 1 mol $L^{-1}$ $LiPF_6$ in EC/PC/EMC (0.14/0.18/0.68 wt.%) have previously been observed by Borong W. et al.[19] The addition of fluoroethylene carbonate (FEC) to the electrolyte was observed to maintain layer continuity while affecting the SEI morphology and the impedance associated with the SEI layer. Contrary to this, heterogeneous formation of electrolyte decomposition compounds such as $Li_2CO_3$ and LiF was observed on LiFePO4 in 1.2 mol $L^{-1}$ $LiPF_6$ in EC:DMC (1:1 by weight) electrolyte by C. C. Chang et. al.[20] It was noted that the addition of tris(pentafluorophenyl) borane (TPFPB) suppress this heterogeneous formation. Similarly, heterogeneous deposition of surface compounds was observed on $LiFePO_4$ using 1 mol $L^{-1}$ LiPF6 in EC:DEC (2:1 by weight) as electrolyte.[21,22]

An interesting complementary X-ray photoelectron spectroscopy (XPS) and *in-situ* atomic force microscopy (AFM) study suggested SEI layers formed on a Highly Oriented Pyrolytic Graphite (HOPG) electrode in 1 mol $L^{-1}$ LiPF6 in EC:DMC (1:1 by weight) consist of a thin and scattered top layer with a dense and more continuous bottom layer.[23] Importantly, the layer formation was observed to dynamically depend on the anodic/cathodic electrode operation. In other words the electrolyte composition, the use of additives and the electrode charge/discharge history was shown to be important for the morphology of decomposition compounds and SEI layer formation on $LiFePO_4$ surfaces. This indicates that the layers observed in Fig. 2(i) could include the primary coating layer and different SEI layers formed when the cathode was stored and cycled.

*4.2 CB Crystallinity*

Relative to the SAED profile from the fresh cathode [Fig. 3], a reduction of peak intensity and peak broadening is observed for both the stored and in particular the aged sample. Decreasing CB crystallinity during storage has previously been observed,[35] and the peak broadening and peak intensity decrease indicates that the CB crystallinity decreases during storage in electrolyte and during cycling. It should be noted that in the aged cathode, the amorphous Si resin - to some extend - tends to smear out the electron diffraction pattern thereby causing additional peak intensity reduction and peak broadening.

*4.3 CB Agglomeration*

The area within the blue dotted circle in Fig. 2(c) shows an aggregate of particles. Here it is important to distinguish between aggregation (gathering of carbon particles) and agglomeration (nucleation of new amorphous phases).[13] A zoom on the aggregated particles is presented in the upper left part of the BF TEM image [Fig. S3(a)]. Here it is seen that the aggregated particles have partly nucleated to form an agglomerate. The same area is shown in the STEM HAADF image in Fig. S3(b). Here the carbon (black), Si-resin (grey) and the LFP (white) can be distinguished by brightness contrast which confirms that the agglomerate primarily consists of carbon. The area shown in Fig. S3 is part of the larger area mapped in Fig. S2 (upper, right part). Fig. S2(f) shows that the Si-resin contains more carbon than the LFP particles and that the agglomerate contain more carbon than the Si-resin.

It should be noted that TEM images only show smaller parts of the cathodes and are therefore prone to severe statistical errors. SEM images [Fig. S4(a) and (b)] show larger parts of the reference and aged electrode with improved statistical information. Comparing the two SEM images, it is observed that relative to the reference electrode a more heterogeneous structure and larger agglomerates are observed in the aged cathode.

It has previously been suggested that Brownian motion may cause aggregation of CB particles.[36,37] The CB aggregation is most likely enhanced by mechanical stress, due to expansion/contraction of LFP upon cycling and/or Fe dissolution from LFP grains.[38] Self-agglomeration of $C_{65}$ and Super P is previously observed in $LiCoO_2$ cathode systems.[39,29] Similarly $C_{65}$ agglomeration has been observed in cycled LMNO/$C_{65}$ electrodes where Fourier transform infrared spectroscopy (FTIR) analysis of the cycled electrode indicated decomposition reactions and possible formation of alkyl carbonates.[40] Thus, in the present study, the observed decrease of CB crystallinity is likely related to the formation of carbon agglomerates.

It should be noted that pristine CB consist of particles that have grown into each other as part of the combustion process so that they share graphitic sheets. This means that the pristine CB – to some extent – already forms agglomerates. However, the agglomerates observed in the aged sample [Fig. S4(b)] are significantly larger than the agglomerates in the pristine CB. Specifically, quasi-crystalline round-shaped CB nanoparticles are observed in both the fresh and the stored cathode, whereas spherical CB nanoparticles are difficult to find in the TEM specimen from the aged sample. Here the carbon has primarily formed amorphous chains rather than spherical particles.

Relative to commercial cathodes, the investigated laboratory-made cathode has a rather poor particle packing with large open pores which enhances carbon particle aggregation and subsequent formation of large agglomerates.[14] Amorphous carbon has a lower electric conductivity than the quasi-crystalline CB,[8,41,42] which means that the formation of large carbon agglomerates with decreased crystallinity along with the increased heterogeneity in the aged sample will likely decrease the electric conductivity in the carbon network.

*4.4 Charge-Discharge Cycling*

The capacity loss observed in $LiFePO_4$ batteries is normally ascribed to the negative graphite electrode,[43] and related to the thick SEI layers of several hundredth of nm that can be formed here.[44] However the thickness of the SEI layers formed on the $LiFePO_4$ particles in this study is much smaller [Fig. 2(f,i), ~30 nm] than the thickness of the SEI layers normally formed at the negative graphite electrode. For this reason, the SEI layer formation at the LiFePO$_4$ surface is not expected to contribute significantly to the observed capacity fade seen for the aged LFP electrode in Fig. 4.

In commercial batteries the resistance of the LiFePO$_4$ electrode is normally significantly larger than the resistance at the graphite electrode.[45] Thus for such batteries SEI layer formation at the LiFePO$_4$ is likely to affect the battery resistance rather than the battery capacity. The DC resistance (before onset of concentration polarization) is fairly constant [Fig. 4] around 2000 Ω cm$^2$ (= 37 mV/40 μA · π· 9 mm$^2$) so the SEI layers formed on the LFP particles during storage and cycling does not seem to affect the DC resistance.

At the low C-rate used in this manuscript the DC over-potentials are related to the critical over-potential required to initiate lithiation/delithiation in the LFP particles.[46] It is believed that this over-potential is not affected by the local increase in current density with capacity fading at the individual LFP particles, since the applied current densities are relatively small – even after the observed capacity fade.

*4.5 Impedance Analysis*

The semi-circle in Fig. 5(a) occurring between 10 kHz and 100 Hz [Fig. 5(b)] is independent of SOC. This is in agreement with the conventional assumption that this arc is related to the interface between the current collector and the cathode materials.[47–51] The mid-low frequency range (between 10 Hz and 10 kHz) [Fig. 5(b)] is characterized by a larger semicircle [Fig. 5(a)] that transitions into a low-frequency tail with an angle of 45° in the Nyquist plot [Fig. 5(a)], which is usually related to respectively electrode/electrolyte interface reactions and diffusion of lithium ions in the LFP particles.[52,53] This part of the spectra is seen to be dependent of SOC in particular at low and high SOC.

Fig. 5(c) and (d) (zoom in (e) and (f)) shows that both the high-frequency part and the low frequency part increase due to cycling. Relative to the reference cathode the aged cathode exhibited increased heterogeneity in the carbon network, carbon agglomeration and decreased carbon crystallinity. Additionally, loss of electric percolation has been reported for the aged LFP CB electrode using low-kV FIB/SEM.[11,14] These degradation mechanisms are expected to decrease the electric conductivity in the CB network thereby increasing the size of the high-frequency arc.

The resistance related to the low-frequency part of the impedance spectra in Fig. 5 is also observed to increase with cycling. The value of the real part of the lowest-frequency impedance in part (c) is larger than the ~2000 $\Omega$ cm$^2$ DC resistance. This is because the impedance is measured at 0% SOC where the overvoltage deviates from the DC overvoltage of 37 mV.

It is important to note that the resistance associated with the high-frequency arc is in the order of 75-175 $\Omega$ cm$^2$ which is small relative to the DC resistance of ~2000 $\Omega$ cm$^2$. Thus the increase with cycling of the size of the high-frequency arc doesn't significantly affect the DC resistance.

## 5. Conclusions

Laboratory assembled electrodes of LiFePO$_4$-Carbon Black (LFP-CB) were investigated at different preparation, storage and cycling stages by complementary TEM and SEM microscopy techniques, charge-discharge capacity curves and impedance spectroscopy.

Decreased crystallinity of CB particles was observed in an electrode cycled 100 times. Additionally agglomeration of carbon was seen in the electrode. The decrease in the crystallinity and CB agglomeration is expected to decrease the electrical conductivity in the CB network. Impedance spectroscopy was performed on the cycled electrode every 10 cycles. The spectra showed an increasing resistance with cycling in the high-frequency impedance arc. This arc was independent of SOC and associated with the CB network – current collector interface. The increase in the arc resistance was linked to a decrease in the conductivity / percolation of the CB network.

Storage in electrolyte and charge/discharge cycling was shown to increase the thickness of amorphous SEI layers formed at the LFP surfaces. Interestingly a multi-layered SEI was formed on the cycled electrode. This suggests that the operational history of the electrode to some extent is recorded by the SEI layer and points towards research that can "read" this history. More systematic studies of the correlation between SEI layer formation and changes in the electrode impedance are required to quantify and solidify these conclusions.

**Acknowledgements**

The authors gratefully acknowledge financial support from the Danish Strategic Research Council through the project "Advanced Lifetime Predictions of Battery Energy Storage" (contract no. 0603-00589B).

**References**

1. Whittingham, M. S. Lithium batteries and cathode materials. *Chem. Rev.* **104,** 4271–4301 (2004).
2. Wang, J. & Sun, X. Olivine LiFePO4: the remaining challenges for future energy storage. *Energy Environ. Sci.* **8,** 1110 (2015).
3. Padhi, A. K., Nanjundaswamy, K. S. & Goodenough, J. B. Phospho-olivines as Positive-Electrode Materials for Rechargeable Lithium Batteries. *J. Electrochem. Soc.* **144,** 1188 (1997).
4. Yamada, A., Chung, S. C. & Hinokuma, K. Optimized LiFePO4 for Lithium Battery Cathodes. *J. Electrochem. Soc.* **148,** A224 (2001).
5. Niu, J. *et al.* In situ observation of random solid solution zone in LiFePO4 electrode. *Nano Lett.* **14,** 4005–4010 (2014).
6. Liao, X.-Z. *et al.* LiFePO4 Synthesis Routes for Enhanced Electrochemical Performance. *J. Electrochem. Soc.* **152,** A231 (2005).
7. Gao, H. *et al.* High rate capability of Co-doped LiFePO4/C. *Electrochim. Acta* **97,** 143–149 (2013).
8. Spahr, M. E., Goers, D., Leone, A., Stallone, S. & Grivei, E. Development of carbon conductive additives for advanced lithium ion batteries. *J. Power Sources* **196,** 3404 (2011).
9. Liu, P. *et al.* Aging Mechanisms of LiFePO4 Batteries Deduced by Electrochemical and Structural Analyses. *J. Electrochem. Soc.* **157,** A499 (2010).
10. Wang, J. & Sun, X. Olivine LiFePO 4 : the remaining challenges for future energy storage. *Energy Environ. Sci.* **8,** 1110–1138 (2015).
11. Scipioni, R. *et al.* Degradation Studies on LiFePO 4 Cathode. *ECS Trans.* **64,** 97 (2015).
12. Roberts, M. R. *et al.* Direct Observation of Active Material Concentration Gradients and Crystallinity Breakdown in LiFePO4 Electrodes During Charge / Discharge Cycling of Lithium Batteries. *J. Phys. Chem. C* **118,** 6548–6557 (2014).
13. Nichols, G. *et al.* A review of the terms agglomerate and aggregate with a recommendation for nomenclature used in powder and particle characterization. *J. Pharm. Sci.* **91,** 2103–2109 (2002).
14. Scipioni, R. *et al.* Electron microscopy investigations of changes in morphology and conductivity of LiFePO4/C electrodes. *J. Power Sources* **307,** 259–269 (2016).
15. Wu, J. *et al.* In situ Raman spectroscopy of LiFePO4: size and morphology dependence during charge and self-discharge. *Nanotechnology* **24,** 424009 (2013).
16. Sacci, R. L. *et al.* Direct visualization of initial SEI morphology and growth kinetics during lithium deposition by in situ electrochemical transmission electron microscopy. *Chem. Commun.* **50,** 2104 (2014).

17. Sugar, J. D. *et al.* High-resolution chemical analysis on cycled LiFePO4 battery electrodes using energy-filtered transmission electron microscopy. *J. Power Sources* **246,** 512 (2014).
18. Samadani, E., Mastali, M., Farhad, S., Fraser, R. A. & Fowler, M. Li-ion battery performance and degradation in electric vehicles under different usage scenarios. *Int. J. Energy Res.* (2015). doi:10.1002/er.3378
19. Wu, B. *et al.* Enhanced electrochemical performance of LiFePO4 cathode with the addition of fluoroethylene carbonate in electrolyte. *J. Solid State Electrochem.* **17,** 811–816 (2013).
20. Chang, C. C. & Chen, T. K. Tris(pentafluorophenyl) borane as an electrolyte additive for LiFePO4 battery. *J. Power Sources* **193,** 834–840 (2009).
21. Herstedt, M. *et al.* Surface Chemistry of Carbon-Treated LiFePO4 Particles for Li-Ion Battery Cathodes Studied by PES. *Electrochem. Solid-State Lett.* **6,** A202 (2003).
22. Stjerdmdahl, M. Stability Phenomena in Novel Electrode Materials for Lithium-ion Batteries. *Acta Universitatis Upsaliensis* (Uppsala University, Sweden, 2007).
23. Shen, C., Wang, S., Jin, Y. & Han, W. Q. In Situ AFM Imaging of Solid Electrolyte Interfaces on HOPG with Ethylene Carbonate and Fluoroethylene Carbonate-Based Electrolytes. *ACS Appl. Mater. Interfaces* **7,** 25441–25447 (2015).
24. William, D. B. & Carter, C. B. *Transmission Electron Microscopy: A Textbook for Materials Science*. (Springer, 2006).
25. Ugarte, D. Curling and closure of graphitic networks under electron-beam irradiation. *Nature* **359,** 707 (1992).
26. Vander Wal, R. L., Tomasek, A. J., Pamphlet, M. I., Taylor, C. D. & Thompson, W. K. Analysis of HRTEM images for carbon nanostructure quantification. *J. Nanoparticle Res.* **6,** 555 (2004).
27. Belenkov, E. a. Formation of Graphite structure in Carbon crystallites. *Inorg. Mater.* **37,** 928–934 (2001).
28. Bi, H., Kou, K. C., Ostrikov, K. & Zhang, J. Q. Graphitization of nanocrystalline carbon microcoils synthesized by catalytic chemical vapor deposition. *J. Appl. Phys.* **104,** 1–7 (2008).
29. Kwon, N. H. The effect of carbon morphology on the LiCoO2 cathode of lithium ion batteries. *Solid State Sci.* **21,** 59–65 (2013).
30. Song, G. M., Wu, Y., Liu, G. & Xu, Q. Influence of AlF3 coating on the electrochemical properties of LiFePO4/graphite Li-ion batteries. *J. Alloys Compd.* **487,** 214–217 (2009).
31. Xu, B., Qian, D., Wang, Z. & Meng, Y. S. Recent progress in cathode materials research for advanced lithium ion batteries. *Mater. Sci. Eng. R Reports* **73,** 51–65 (2012).
32. Lawder, M. T., Northrop, P. W. C. & Subramanian, V. R. Model-Based SEI Layer Growth and Capacity Fade Analysis for EV and PHEV Batteries and Drive

Cycles. *J. Electrochem. Soc.* **161,** A2099–A2108 (2014).

33. Li, D. *et al.* Modeling the SEI-Formation on Graphite Electrodes in LiFePO4 Batteries. *J. Electrochem. Soc.* **162,** A858–A869 (2015).

34. Yamamoto, K. *et al.* Dynamic visualization of the electric potential in an all-solid-state rechargeable lithium battery. *Angew. Chemie - Int. Ed.* **49,** 4414–4417 (2010).

35. Younesi, R. *et al.* Analysis of the Interphase on Carbon Black Formed in High Voltage Batteries. *J. Electrochem. Soc.* **162,** A1289–A1296 (2015).

36. Zhu, M., Park, J. & Sastry, A. M. Particle Interaction and Aggregation in Cathode Material of Li-Ion Batteries: A Numerical Study. *J. Electrochem. Soc.* **158,** A1155 (2011).

37. Li, J., Armstrong, B. L., Daniel, C., Kiggans, J. & Wood, D. L. Optimization of multicomponent aqueous suspensions of lithium iron phosphate (LiFePO4) nanoparticles and carbon black for lithium-ion battery cathodes. *J. Colloid Interface Sci.* **405,** 118–124 (2013).

38. Zhi, X. *et al.* The cycling performance of LiFePO4/C cathode materials. *J. Power Sources* **189,** 779–782 (2009).

39. Hong, J. K., Lee, J. H. & Oh, S. M. Effect of carbon additive on electrochemical performance of LiCoO2 composite cathodes. *J. Power Sources* **111,** 90–96 (2002).

40. Arbizzani, C., Da Col, L., De Giorgio, F., Mastragostino, M. & Soavi, F. Reduced Graphene Oxide in Cathode Formulations Based on LiNi0.5Mn1.5O4. *J. Electrochem. Soc.* **162,** A2174–A2179 (2015).

41. Chu, P. K. & Li, L. Characterization of amorphous and nanocrystalline carbon films. *Mater. Chemi* **96,** 253–277 (2006).

42. Pandolfo, A. G. & Hollenkamp, A. F. Carbon properties and their role in supercapacitors. *J. Power Sources* **157,** 11–27 (2006).

43. Klett, M. *et al.* Non-uniform aging of cycled commercial LiFePO4//graphite cylindrical cells revealed by post-mortem analysis. *J. Power Sources* **257,** 126–137 (2014).

44. Zhang, H. L., Li, F., Liu, C., Tan, J. & Cheng, H. M. New insight into the solid electrolyte interphase with use of a focused ion beam. *J. Phys. Chem. B* **109,** 22205–22211 (2005).

45. Belt, J. R., Bernardi, D. M. & Utgikar, V. Development and Use of a Lithium-Metal Reference Electrode in Aging Studies of Lithium-Ion Batteries. *J. Electrochem. Soc.* **161,** A1116–A1126 (2014).

46. Cogswell, D. A. & Bazant, M. Z. Theory of coherent nucleation in phase-separating nanoparticles. *Nano Lett.* **13,** 3036–3041 (2013).

47. Zhu, Y., Xu, Y., Liu, Y., Luo, C. & Wang, C. Comparison of electrochemical performances of olivine NaFePO4 in sodium-ion batteries and olivine LiFePO 4 in lithium-ion batteries. *Nanoscale* **5,** 780–787 (2013).

48. Chang, Y.-C. & Sohn, H.-J. Electrochemical Impedance Analysis for Lithium Ion

Intercalation into Graphitized Carbons. *J. Electrochem. Soc.* **147,** 50 (2000).

49. Liao, X.-Z. *et al.* Electrochemical Behavior of LiFePO4/C Cathode Material for Rechargeable Lithium Batteries. *J. Electrochem. Soc.* **152,** A1969 (2005).

50. Gaberscek, M., Moskon, J., Erjavec, B., Dominko, R. & Jamnik, J. The Importance of Interphase Contacts in Li Ion Electrodes: The Meaning of the High-Frequency Impedance Arc. *Electrochem. Solid-State Lett.* **11,** A170 (2008).

51. Illig, J. *et al.* Separation of Charge Transfer and Contact Resistance in LiFePO4-Cathodes by Impedance Modeling. *J. Electrochem. Soc.* **159,** A952 (2012).

52. Meyers, J. P., Doyle, M., Darling, R. M. & Newman, J. The Impedance Response of a Porous Electrode Composed of Intercalation Particles. *J. Electrochem. Soc.* **147,** 2930 (2000).

53. Gao, F. & Tang, Z. Kinetic behavior of LiFePO4/C cathode material for lithium-ion batteries. *Electrochim. Acta* **53,** 5071 (2008).

**TABLES**

Table 1: Overview of the investigated LFP-CB samples

| Number | Name | Description | Sample |
|---|---|---|---|
| 1 | pristine CB | as-received CB powder | Powder (TEM) |
| 2 | pristine LFP | as-received LFP powder | Powder (TEM) |
| 3 | fresh cathode | LFP-CB (+ binder) | Powder (TEM) |
| 4 | stored cathode | sample 3 stored in electrolyte for 72 h | Powder (TEM) |
| 5 | reference cathode | after 2 charge/discharge cycles | FIB/SEM |
| 6 | aged cathode | after 100 charge/discharge cycles | Lamellar (TEM) + FIB/SEM |

Table 2: SAED Peak data

| Diffraction quantity | (002) peak | | | (101) peak | | |
|---|---|---|---|---|---|---|
| | Fresh | Stored | Aged | Fresh | Stored | Aged |
| Intensity (a.u.) | 1.13±0.02 | 0.96±0.02 | 0.15±0.02 | 0.47±0.02 | 0.36±0.02 | 0.30±0.02 |
| FWHM (a.u.) | 0.22±0.02 | 0.25±0.02 | 0.35±0.02 | 0.36±0.02 | 0.47±0.02 | 0.58±0.02 |

**FIGURES**

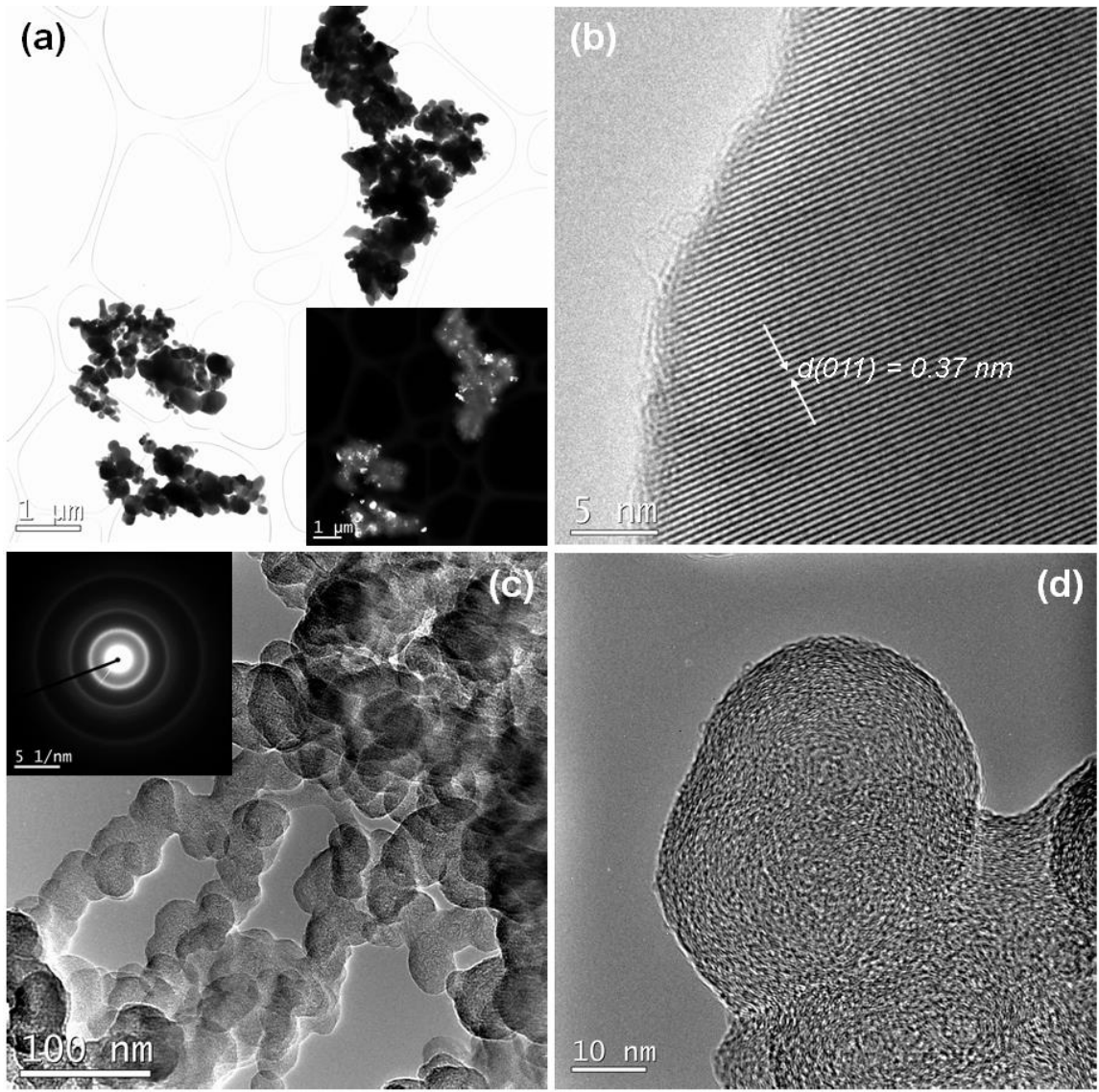

FIG 1. (a) Bright-field TEM image of pristine (off-the-shelf) LFP nanoparticles on lacey carbon grid, the inset shows a (200)-reflected dark-field image of the corresponding area; (b) HRTEM image of a single-crystal LFP nanoparticle; (c) Bright-field TEM image of pristine (off-the-shelf) CB nanoparticles. The inset shows a selected area electron diffraction pattern of the corresponding CB nanoparticles; (d) HRTEM image of a typical CB nanoparticle.

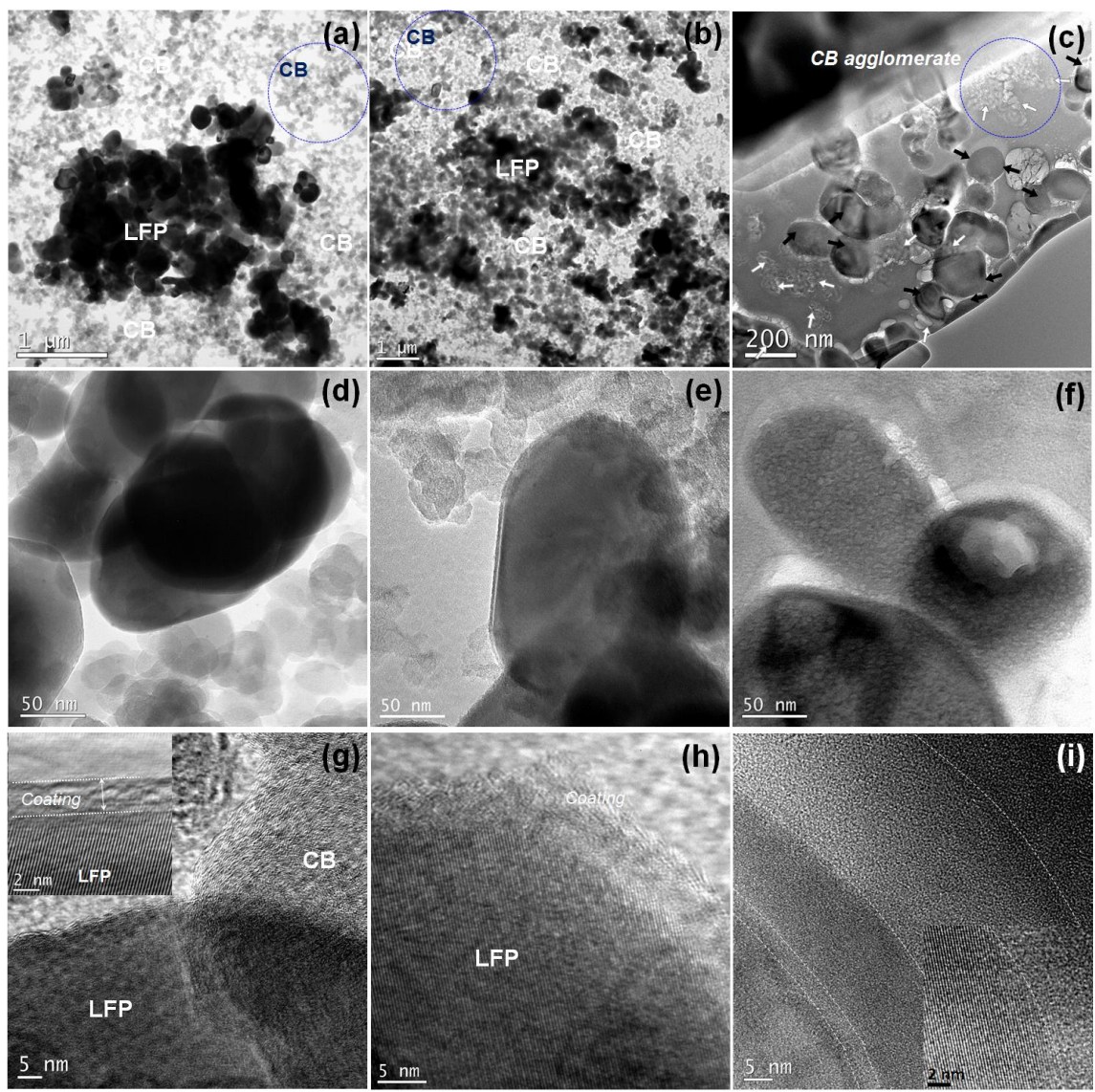

FIG. 2. TEM images of the (a,d,g) fresh (b,e,h) stored and (c,f,i) aged LFP-CB cathode. The inset in (g) presents the magnified primary coating layer on LFP nanoparticles, and inset in (i) shows the magnified LFP/SEI layer interface.

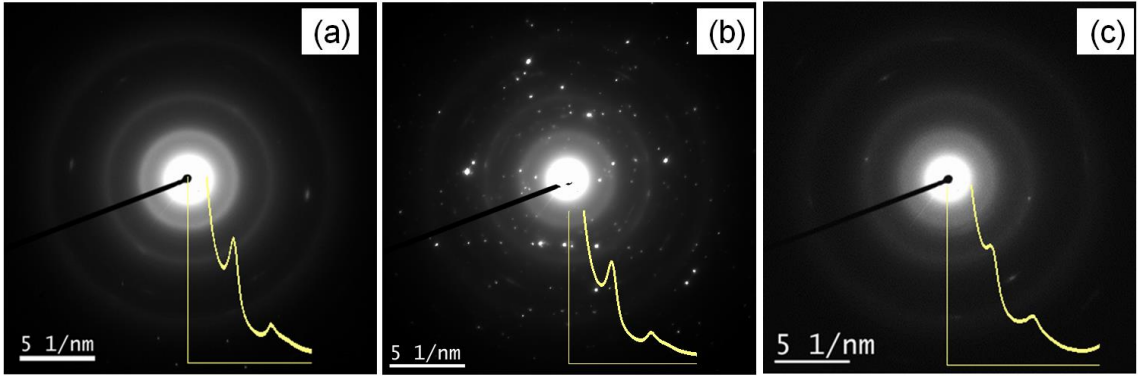

FIG. 3. SAED electron diffraction patterns of CB particles for: (a) fresh cathode, (b) stored cathode and (c) aged cathode recorded inside the dotted blue rings in Fig. 2(a), Fig. 2(b) and Fig. 2(c) respectively. The insets show normalized intensity profiles.

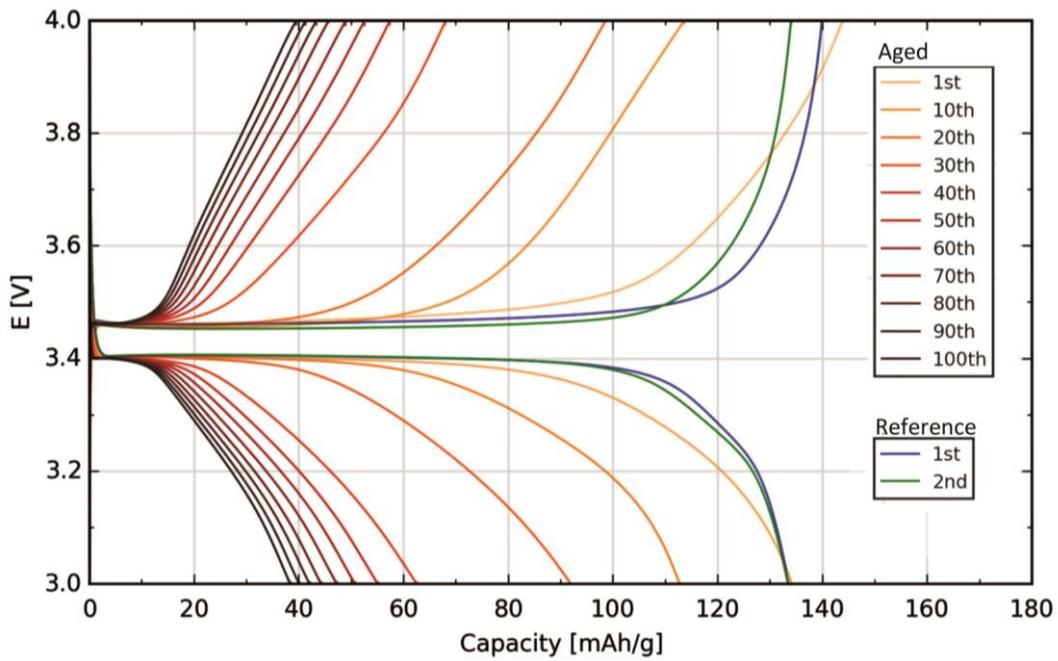

FIG. 4. Charge/discharge curves for the reference and aged cathodes recorded at 0.1 C.

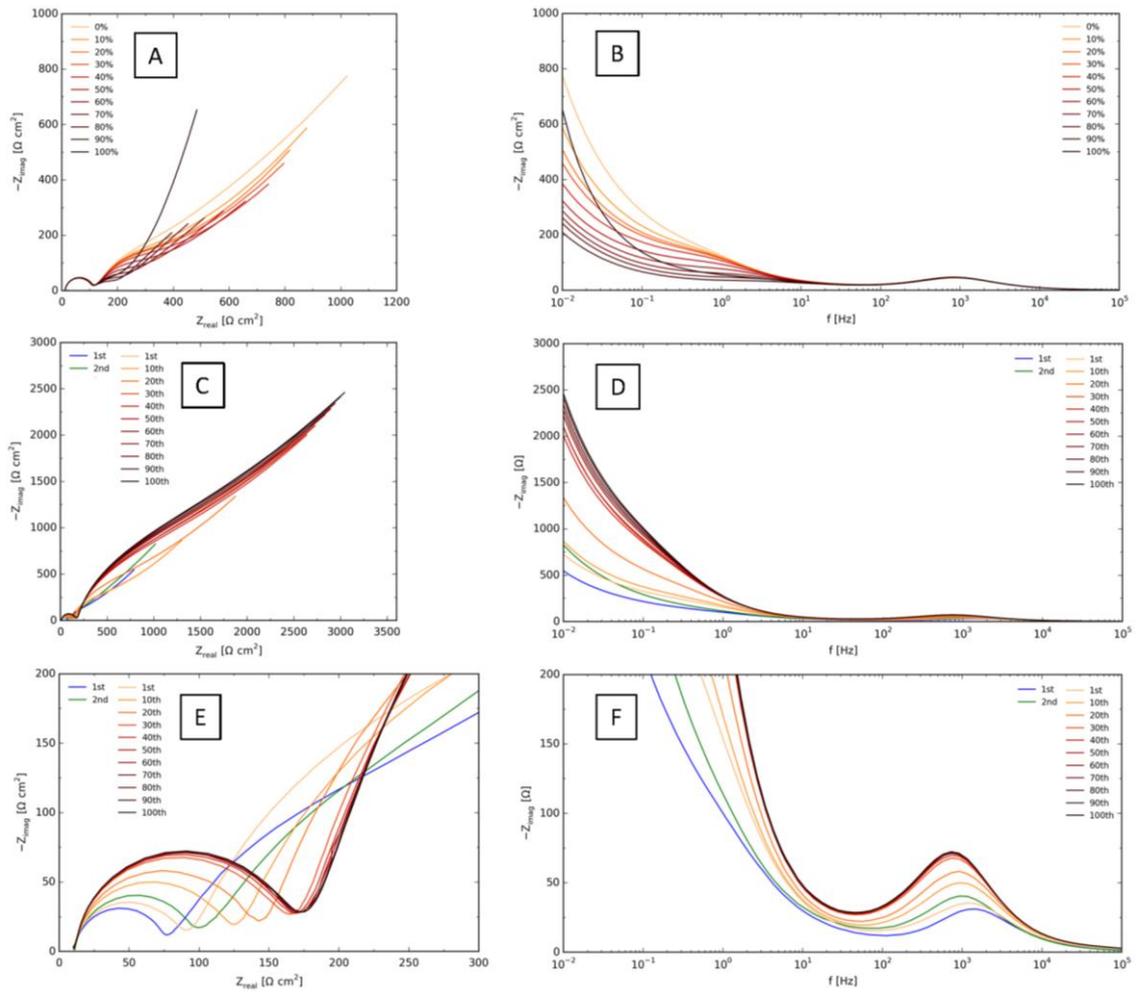

FIG. 5. (a) Nyquist and (b) Bode plots of impedance spectra recorded at various SOC with the aged electrode after 5 cycles. (c) Nyquist and (d) Bode plots of impedance spectra recorded with the reference electrode (blue and green spectra), and with the aged electrode (yellow-red-black spectra) (e) Nyquist and (f) Bode plots showing a zoom of the high-frequency part of the spectra presented in (c) and (d).

**SUPPLEMENTARY DATA**

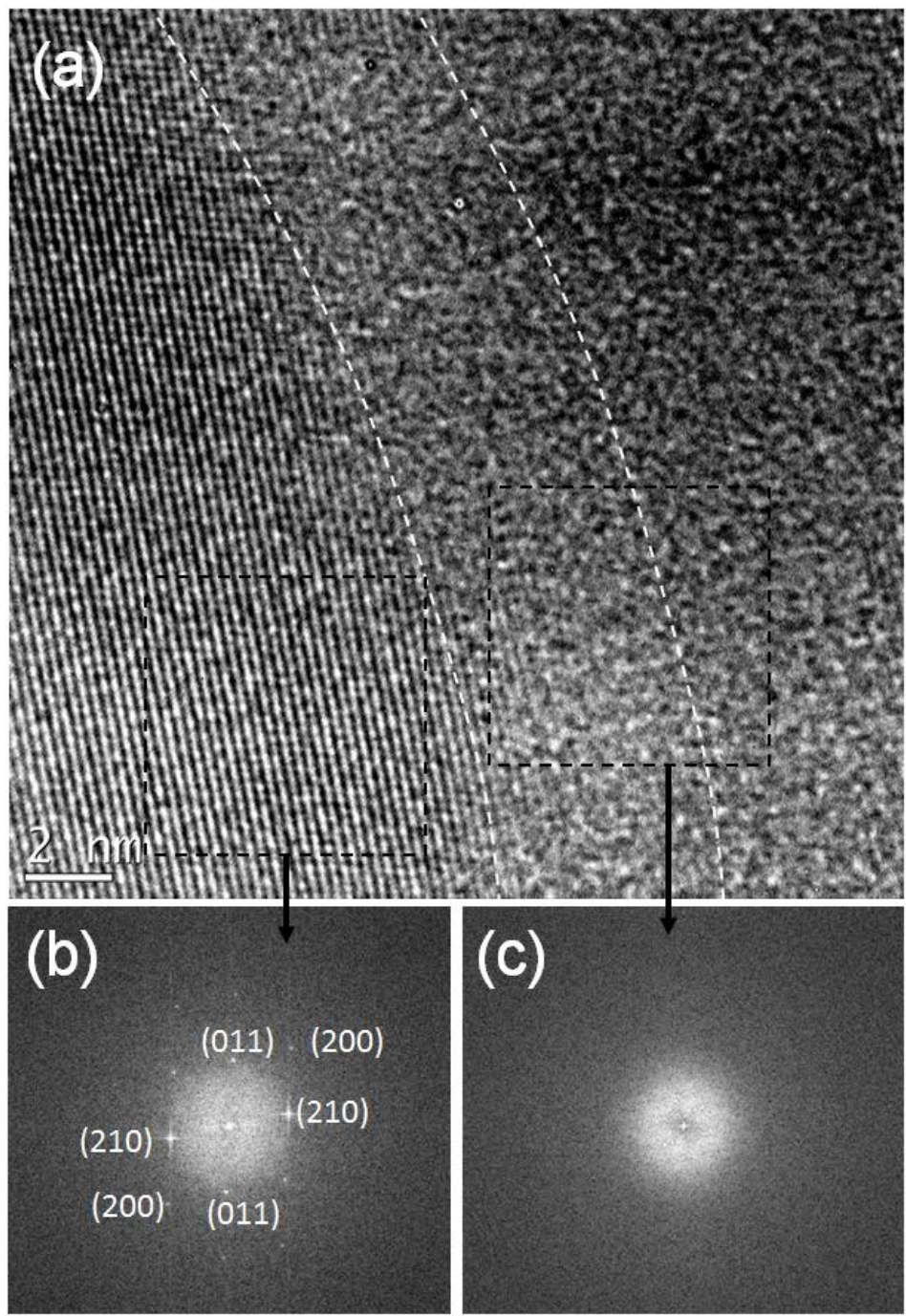

Fig. S1. (a) HRTEM image of aged LFP-CB showing the LFP/SEI layer interfaces. (b) Fast Fourier transform (FFT) patterns from the LFP dotted area. (c) FFT of the SEI layer dotted area. The crystalline structure of the LFP particle and the amorphous structure of the SEI layer are observed.

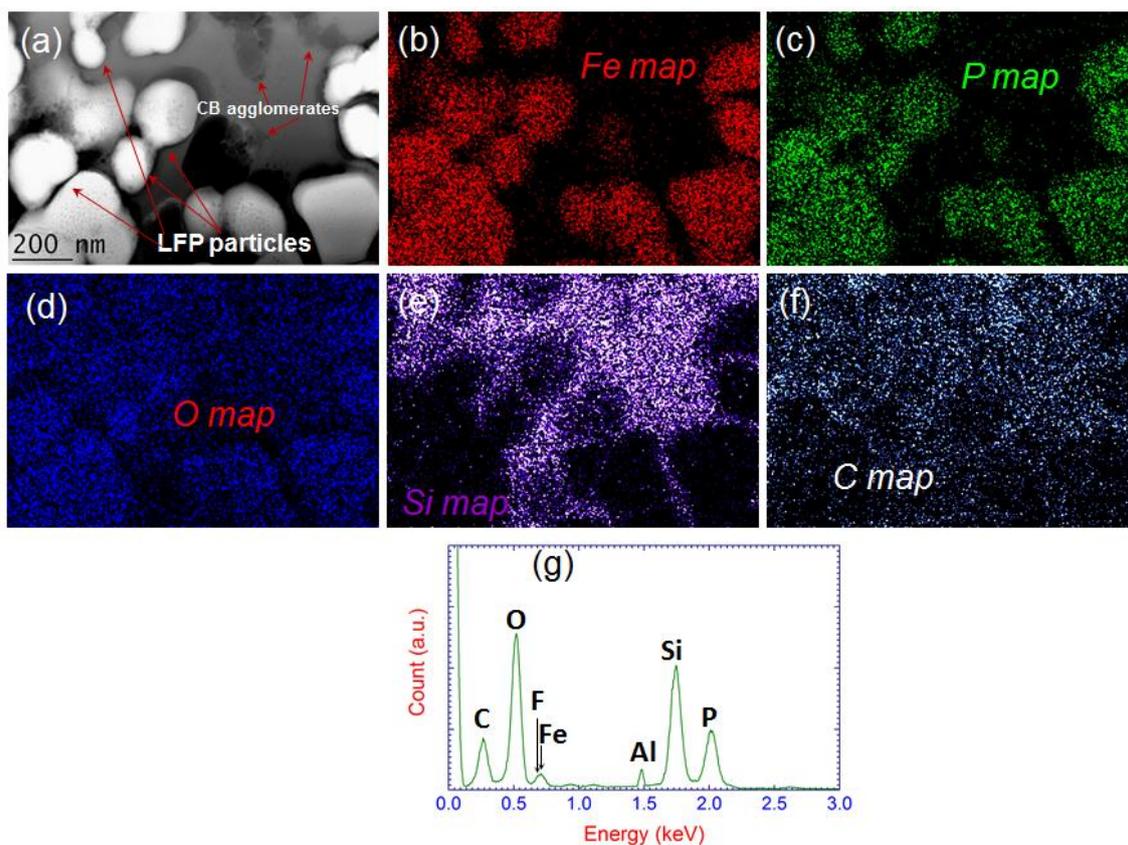

Fig. S2. (a) STEM-HAADF image of the LFP-CB aged cathode. (b-f) STEM-EDX elementary maps of Fe, P, O, Si and C in the aged cathode from the HAADF image. (g) The average EDX spectrum of the map.

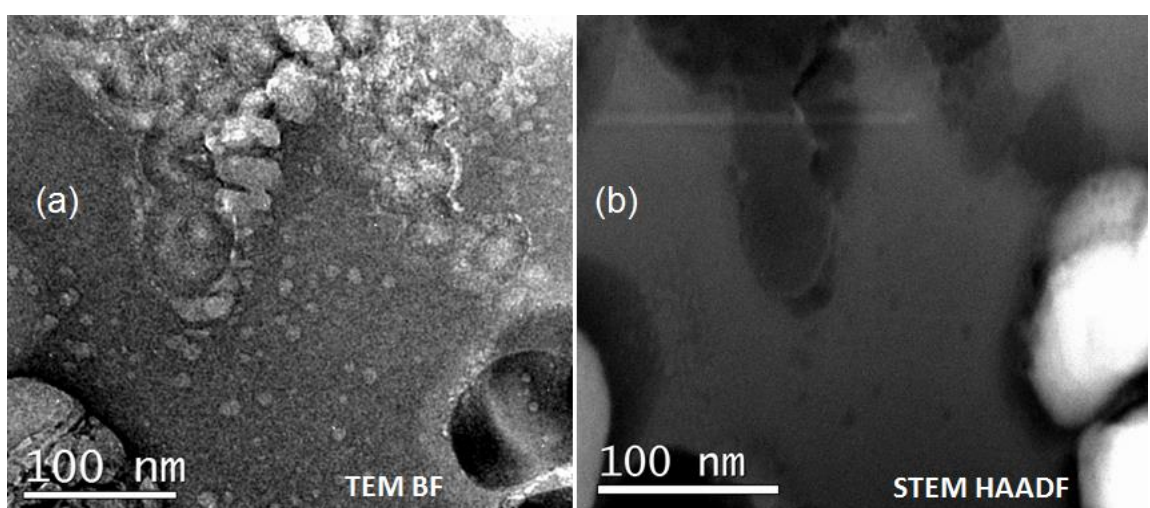

Fig. S3. (a) A BF-TEM image zoom of the carbon agglomerate shown in the blue dotted circle in Fig. 2(c), (b) a STEM HAADF image of the area in (a).

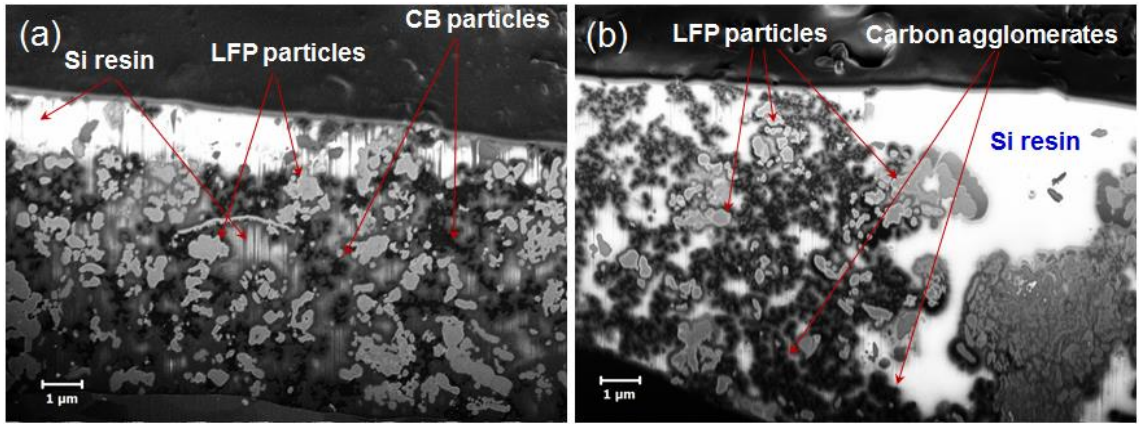

Fig. S4. *In-lens* FIB/SEM images of the reference (a) and aged (b) cathode. Charging during image recording increases the Si resin brightness in some areas in the fresh sample and almost everywhere in the aged sample.